\documentclass[10pt,journal,compsoc]{IEEEtran}

\ifCLASSOPTIONcompsoc
  \usepackage[nocompress]{cite}
\else
  \usepackage{cite}
\fi
\ifCLASSINFOpdf
\usepackage[pdftex]{graphicx}
\else
\usepackage[dvips]{graphicx}
\fi
\usepackage{array}
\usepackage[hyphens]{url}
\usepackage{color, colortbl}

\usepackage[shortlabels]{enumitem}

\usepackage{multirow}
\usepackage{booktabs}

\usepackage{hyperref}

\begin{document}

\newcommand{\ggplottwo}{ggplot2}

%
\title{Effectiveness of Area-to-Value Legends and Grid Lines in Contiguous Area Cartograms}
%
%
%
%

\author{Kelvin~L.~T.~Fung, Simon~T.~Perrault, and Michael~T.~Gastner
  \IEEEcompsocitemizethanks{
    \IEEEcompsocthanksitem{
      K.~L.~T.~Fung is with Yale-NUS College, Singapore 138527, and University College London, WC1E 6BT.
    }
    \IEEEcompsocthanksitem{
    M.~T.~Gastner is with the Singapore Institute of Technology, Singapore 138683.
    }
    \IEEEcompsocthanksitem{
      S.~T.~Perrault is with the Singapore University of Technology
      and Design, Singapore 487372.
    }
  }
}

%
%

\markboth{IEEE Transactions on Visualization and Computer Graphics}%
{Fung \MakeLowercase{\textit{et al.}}: }
%



\IEEEtitleabstractindextext{%
\begin{abstract}

A contiguous area cartogram is a geographic map in which the area of each region is proportional to numerical data (e.g., population size) while keeping neighboring regions connected.
In this study, we investigated whether value-to-area legends (square symbols next to the values represented by the squares' areas) and grid lines aid map readers in making better area judgments.
We conducted an experiment to determine the accuracy, speed, and confidence with which readers infer numerical data values for the mapped regions.
We found that, when only informed about the total numerical value represented by the whole cartogram without any legend, the distribution of estimates for individual regions was centered near the true value with substantial spread.
Legends with grid lines significantly reduced the spread but led to a tendency to underestimate the values.
Comparing differences between regions or between cartograms revealed that legends and grid lines slowed the estimation without improving accuracy.
However, participants were more likely to complete the tasks when legends and grid lines were present, particularly when the area units represented by these features could be interactively selected.
We recommend considering the cartogram's use case and purpose before deciding whether to include grid lines or an interactive legend.
\end{abstract}

\begin{IEEEkeywords}
  Cartogram,
  geovisualization,
  interactive data exploration,
  quantitative evaluation
\end{IEEEkeywords}}

\maketitle

\IEEEdisplaynontitleabstractindextext

%
\IEEEpeerreviewmaketitle

\ifCLASSOPTIONcompsoc
\IEEEraisesectionheading{\section{Introduction}\label{sec:introduction}}
\else
\section{Introduction}
\label{sec:introduction}
\fi

\begin{figure}[!ht]
  \centering
    \includegraphics[width=3in]{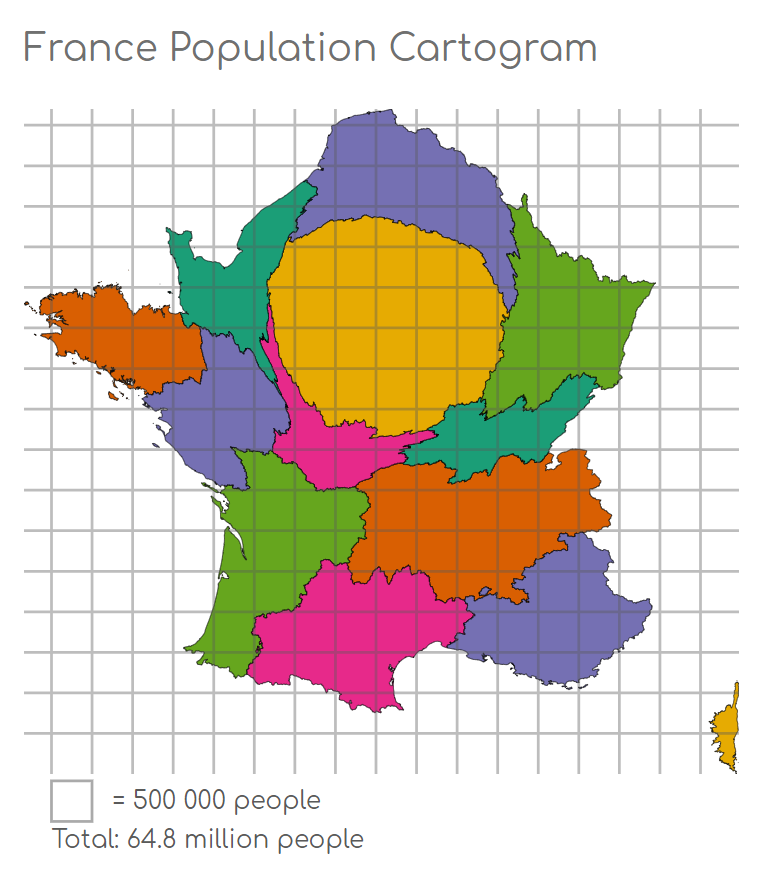}
  \caption{Example of a cartogram that has a static area-to-value legend (square in the bottom left) and grid lines.}
  \label{fig:gridlines}
\end{figure}

%
%
%
%
\IEEEPARstart{A}{s} contemporary computer technology has simplified the production of data visualizations, researchers are now interested in evaluating and improving existing design practices for visualizations displayed on a computer screen.
A cartogram is a type of data visualization for which there are currently only a few design guidelines~\cite{dent_communication_1975, tingsheng_motivating_2020}.
A contiguous area cartogram is a special type of cartogram in which the area of each region is rescaled according to quantitative statistical data without changing the underlying map topology (i.e., neighboring regions must remain connected).
Because contiguous area cartograms can simultaneously visualize geography and statistics, they are used, for example, in newspaper articles~\cite{riedmann_erde_2021}, textbooks~\cite{de_veaux_stats_2015}, and online tutorials~\cite{clark_carbon_2012}.
A large collection of contiguous cartograms is available from the website of the Worldmapper project~\cite{gotthard_worldmapper_2022}, which aims to visualize various global statistics~\cite{hennig_remapping_2010}.
An example of a contiguous area cartogram is shown in Fig.~\ref{fig:gridlines}.
The term ``cartogram'' is also used for many other related map designs (e.g., distance cartograms~\cite{shimizu_new_2009} and non-contiguous area cartograms~\cite{olson_noncontiguous_1976}).
Hereinafter, we refer to contiguous area cartograms simply as ``cartograms'' for the sake of brevity.

Cartograms satisfy Tufte's principle of graphical integrity: ``The representation of numbers, as physically measured on the surface of the graphic itself, should be directly proportional to the numerical quantities represented''~\cite[p.~56]{tufte_visual_2001}.
However, Dent~\cite[p.~164]{dent_communication_1975} reported that users, who generally considered cartograms to be ``innovative'' and ``interesting,'' rated cartograms only as average when asked whether they felt that relative magnitudes were clearly shown.
Hence, Dent recommended including an area-to-value legend in every cartogram to assist map readers with quantitative assessments.
Such a legend comprises a square symbol that shows how the area of the square is to be converted to the numerical value printed next to the square (Fig.~\ref{fig:gridlines}).
We build on Dent's recommendation and propose two additional features: grid lines and a selectable legend (i.e., an interactive area-to-value legend that allows the user to choose the area of the square).
The purpose of this study is to evaluate whether static legends, grid lines, and selectable legends help map readers retrieve quantitative information from cartograms.

\begin{figure*}[!ht]
  \centering
    \includegraphics[width=5.5in]{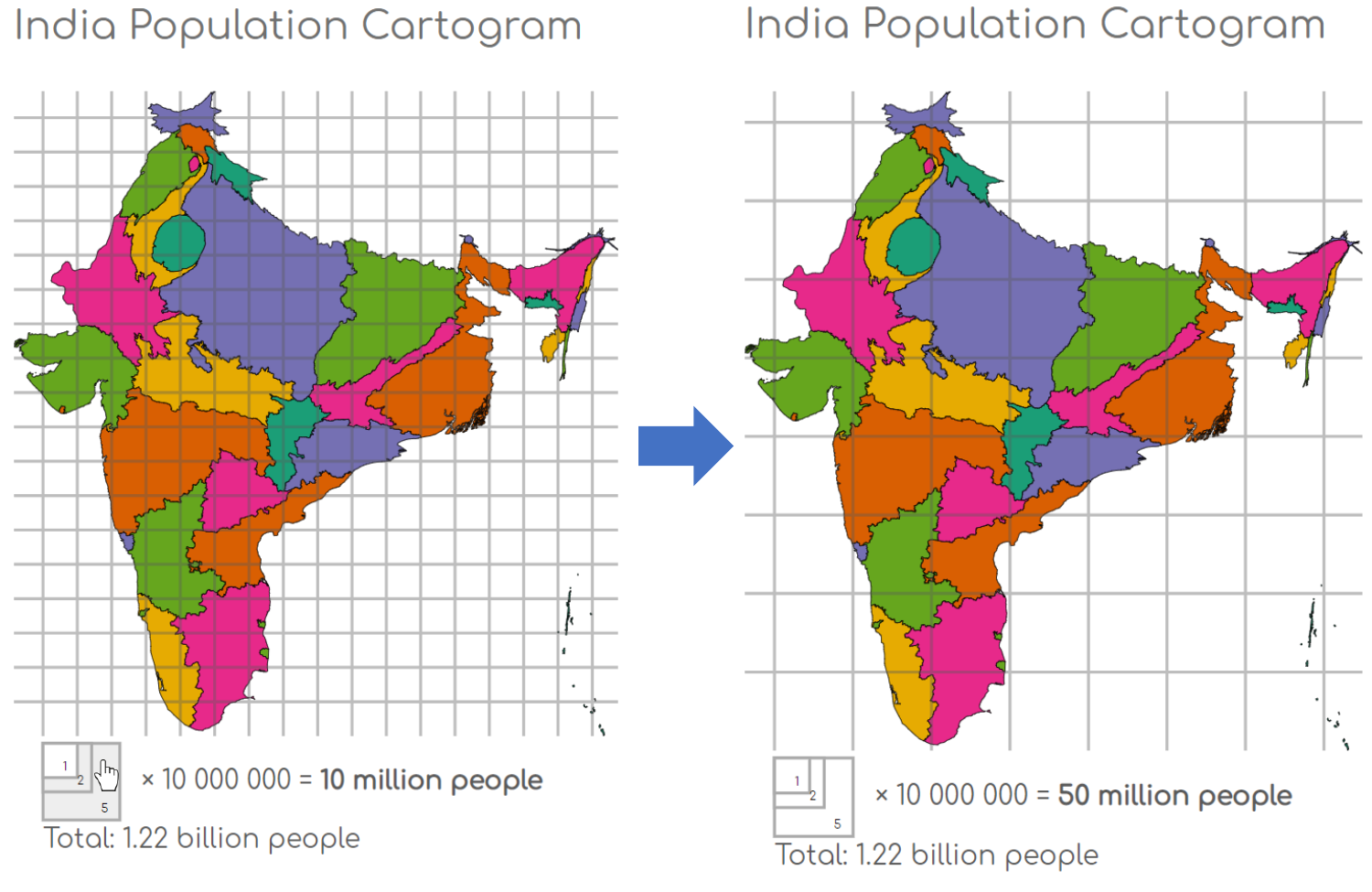}
  \caption{Illustration of the interactive selectable legend feature used in the experiment.
  Participants could select from three squares placed below the cartogram.
  In this example, the areas of the squares corresponded to populations of 10, 20, and 50 million.
  The active square appeared white.
  The other squares had a gray fill color and were stacked below the active square.
  When the participant clicked on a legend square that was not currently selected, the newly selected square was highlighted, and the space between the grid lines either expanded or contracted to correspond to the new legend square size.
  The example shown in this figure demonstrates the effect of switching from a legend square that corresponds to 10 million people (left) to a square that corresponds to 50 million people (right).}
  \label{fig:selectable}
\end{figure*}

\subsection{Features Studied}

\subsubsection{Static Legend}
\label{sec:static_legend}
We use the term ``static legend'' to refer to a single, non-interactive square symbol next to the associated numerical value.
For the size of the legend square, we chose an area that represented approximately $1\%$ of the cartogram's total area, consistent with Dent's recommendation that this area should be at ``the low end of the value range''~\cite[p.~167]{dent_communication_1975} (see Section~\ref{sec:previous_evaluations}). 
We only used values that were ``nice'' numbers (i.e., powers of 10 multiplied by 1, 2, or 5)~\cite{wilkinson_grammar_2012, tingsheng_motivating_2020}.

\subsubsection{Grid Lines}
Grid lines are vertical and horizontal lines overlaid on a cartogram, as shown in Fig.~\ref{fig:gridlines}. 
To ensure that the lines did not obfuscate the underlying map, we used thin, translucent gray lines (line width 2 pixels, hex code \#5A5A5A, $\alpha = 0.4$). 
The size of each square in the grid corresponded to the size of the legend square, whose side length was between 20 pixels and 70 pixels, depending on the data shown.
For comparison, maps and cartograms had, on average, a total width of approximately 500 pixels and a height of 450 pixels.
We aligned the leftmost vertical grid line with the left edge of the legend square.
The position of the legend square was fixed relative to the bounding box of the cartogram as follows:
\begin{itemize}
\item
  In the $x$-direction, the left edge of the legend square was 20 pixels to the right of the left edge of the cartogram's bounding box.
\item
  In the $y$-direction, the top edge of the legend square was 15 pixels below the bottom edge of the cartogram's bounding box.
\end{itemize}

\subsubsection{Selectable Legend with Grid Lines}
In our experiment, a selectable legend consisted of three squares of different sizes overlapping each other, as shown in Fig.~\ref{fig:selectable}. 
We chose to have three squares because Cox~\cite{cox_anchor_1976}  found that having at least three symbol sizes on proportional symbol maps can reduce estimation errors. 
When a user hovered the mouse cursor over a legend square that was not currently selected, the mouse cursor changed into a pointer, indicating that the user could perform a left click. 
Upon left clicking, the newly selected legend square was highlighted, the legend magnitude was updated, and the space between the grid lines either expanded or contracted to correspond to the legend square size.  

For the smallest square, we chose an area that represented approximately $1\%$ of the cartogram's total data value, and we ensured that the value was a ``nice'' number, as described in Section~\ref{sec:static_legend}.
We then computed three consecutive nice numbers, with the value of the smallest square as the smallest number.
The areas of the medium and large squares represented the second-smallest and largest nice numbers respectively. 
For example, in Fig.~\ref{fig:selectable}, the small square represents 10 million people, the medium square represents 20 million people, and the large square represents 50 million people.

\section{Related Work}

\subsection{Grid Lines in Statistical Graphs and Maps}

Despite doubts regarding the general usefulness of grid lines~\cite{few_grid_2005}, many software packages automatically add them to statistical graphs by default (e.g., Excel\textsuperscript \textregistered and \ggplottwo~\cite{wickham_ggplot2_2021}).
An experiment by Heer and Bostock~\cite{heer_crowdsourcing_2010} confirmed that grid lines can indeed be a valuable addition because they improved the accuracy of estimating distances in the direction perpendicular to the lines.
The effect was already significant when only 10 grid lines were shown in the plot.
Increasing the number of grid lines did not lead to further improvements.
The performance even deteriorated with dense grid lines; thus, Heer and Bostock recommended that grid lines be separated by at least 8 pixels.
Further recommendations in the literature include making grid lines thin and light~\cite{boers_designing_2018, schwabish_economists_2014}, gray~\cite{tufte_visual_2001}, and partially transparent~\cite{stone_alpha_2008, heer_crowdsourcing_2010, bartram_whisper_2011}.

World maps are often overlaid with curved grid lines representing a longitude-latitude graticule after applying a map projection to a reference ellipsoid or geoid~\cite{snyder_map_1987}.
For maps of a single country, it is also common to overlay a national grid (e.g., the Ordnance Survey National Grid shown on many maps of regions in the United Kingdom~\cite{ordnance_survey_using_2016}).
Support for this practice comes from human-subject experiments revealing that square grids overlaid on maps improved the recall of point locations on maps~\cite{bestgen_grid_2013}.
Curved grid lines have been added to cartograms for aesthetic reasons unrelated to the visualized data~\cite{woytinsky_world_1953} or to indicate the distortion of the underlying density-equalizing map projection~\cite{tobler_continuous_1973}.
However, we are not aware of previous uses of regular square grids overlaid on contiguous cartograms to enable the visual conversion between area and data value.

\subsection{Previous Evaluations of Cartograms}
\label{sec:previous_evaluations}

Area is the visual variable with which cartograms communicate quantitative data.
According to Cleveland's and McGill's theory of graphical perception~\cite{cleveland_graphical_1984}, humans are generally poorer at judging areas than they are at judging lengths or angles.
The ability of readers to extract quantitative information from areas shown in cartograms has been the focus of previous studies.
In 1975, Dent conducted an experiment to evaluate the communication aspects of cartograms~\cite{dent_communication_1975}. 
Dent reported that participants found cartograms ``confusing and difficult to read, but at the same time [they appeared] interesting, generalized, innovative, unusual, and having---as opposed to lacking---style'' ~\cite[p.~167]{dent_communication_1975}.
From the quantitative data collected during the experiment, Dent concluded that ``better magnitude estimation is achieved when at least a key symbol at the low end of the value range is included''~\cite[p.~160]{dent_communication_1975}.
Dent did not specify what the low end of the value range is, but the sample cartogram he provided has a key that represents approximately $1\%$ of the total population. 
We followed Dent's example by having each legend in our experiment represent approximately $1\%$ of the total value.

Subsequent studies have often pitted cartograms against other types of thematic maps (e.g., choropleth maps and proportional symbol maps). 
Kaspar~\textit{et al.}~\cite{kaspar_empirical_2011} conducted a study that assessed how well map readers made spatial inferences about data presented in cartograms versus data presented in choropleth maps.
They found that cartograms and choropleth maps with graduated circles were equally effective when participants had to answer simple questions.
However, when questions were complex, participants performed better with choropleth maps.
Similar results were also obtained in other cognitive experiments~\cite{aschwanden_kognitionsstudien_1998, sun_effectiveness_2010,han_experimental_2017}.
Recently, Nusrat~\textit{et al.}~\cite{nusrat_recognition_2020} compared contiguous cartograms and Dorling (i.e., circular) cartograms (see~\cite{dukaczewski_application_2020} and~\cite{gastner_cartogram_2022} for classifications of cartogram designs) with respect to recognizability and recall.
They reported that Dorling cartograms led to fewer errors in synoptic tasks (i.e., tasks in which participants had to summarize high-level differences between two cartograms).
However, they found no significant differences between contiguous and Dorling cartograms in terms of recalling specific details of the visualizations.

In a prior study, Nusrat~\textit{et al.}~\cite{nusrat_evaluating_2016} evaluated the effectiveness of four types of area cartograms: contiguous, non-contiguous, rectangular, and Dorling cartograms. 
Their experimental results show that there is partial evidence that contiguous and Dorling cartograms lead to the lowest error rate in the comparison of areas.
In this study, we investigated whether value-to-area legends and grid lines can aid map readers in making more accurate area judgments.

\subsection{Addition of Interactive Features}
In this digital age, features that can be added to cartograms are no longer constrained by the limits of pen and paper. 
As Goodchild noted, cartography's ``true potential lies in less conventional methods of analysis and display and in the degree to which it can escape its traditional constraints''~\cite[p.~311]{goodchild_stepping_1988}. 
An early study conducted in 1999 by Peterson~\cite{peterson_active_1999} explored the potential of computer technology by developing an interactive legend that controlled a cartographic animation using JavaScript.
In a more recent study, Duncan~\textit{et al.}~\cite{duncan_task-based_2021} experimentally evaluated whether cartograms can convey information more effectively if they possess three additional interactive features: cartogram-switching animations, linked brushing, and infotips.
They found that these interactive features significantly reduced error rates in synoptic tasks.
The interactive features were implemented using JavaScript and are currently deployed on the web application~\href{https://go-cart.io}{go-cart.io}~\cite{tingsheng_go-cartio_2019, gastner_creating_2021}.
In our study, we also tapped the potential of computer technology.
We considered whether an interactive feature (a selectable legend) allows map readers to retrieve information from cartograms more effectively.

\subsection{Task Types to Evaluate the Effectiveness of Cartograms}

To assess how well cartograms convey information, participants should be asked to perform a variety of task types that encapsulate the different ways cartograms can be used as a source of information. 
To this end, Nusrat and Kobourov designed a task taxonomy for cartograms with ten objective-based task types (i.e., task types that ``focus on user intent, or what the user wishes to perform'')~\cite[p.~62]{nusrat_task_2015}.
These task types are characterized by verbs (e.g., ``Compare'' and ``Identify'') that do not specify the method or the feature that the participant should use when performing the tasks.
Following these examples, we also adopted an objective-based task taxonomy for our experiment to evaluate the effectiveness of legends and grid lines.

\section{Experiment}

\subsection{Task Types}
\label{subsec:tasks}

\subsubsection{Choice of Tasks}
From the objective-based task taxonomy by Nusrat and Kobourov~\cite{nusrat_task_2015}, we selected four task types for which legends and grid lines are conceivably relevant: \textit{Compare}, \textit{Detect Change}, \textit{Cluster}, and \textit{Find Top}.
We did not include tasks of the types \textit{Locate}, \textit{Recognize}, \textit{Identify}, and \textit{Find Adjacency}.
Those tasks are relevant for comparisons among different cartogram designs (e.g., contiguous versus non-contiguous).
However, in our experiment all cartograms were contiguous; thus, no comparison among different designs was needed.
We also excluded Nusrat's and Kobourov's \textit{Filter} task type because of its similarity with \textit{Compare}.
Tasks of both types require participants to decide whether a region is larger or smaller than a reference region; the only difference is that \textit{Compare} tasks can be answered by comparing only two regions, whereas \textit{Filter} tasks need multiple comparisons.

We used the \textit{Compare} and \textit{Detect Change} task types twice: once for administrative units (e.g., states in the USA or provinces in Nepal) and a second time for ``zones,'' which are spatially contiguous areas formed by aggregating administrative units.
We divided the respective countries into two zones, and both zones were approximately equally large.
The zones were colored yellow and purple, respectively, to achieve a clear color contrast.
For example, we considered New Zealand's South Island as a ``zone'' with seven administrative units (Canterbury, Marlborough, Nelson, Otago, Southland, Tasman, and West Coast).
The South Island was colored yellow, and the other zone (i.e., the North Island) was colored purple.
We included task types for zones as well as for individual administrative units so that participants had to make area comparisons on a variety of length scales.
The \textit{Compare Zones} and \textit{Detect Change in Zone} tasks have a similar aim as \textit{Summarize} tasks in Nusrat and Kobourov's task taxonomy; all three task types ask participants to see the ``big picture'' beyond individual administrative units.
We also added a task type that we call \textit{Estimate Administrative Unit}, which is not part of Nusrat and Kobourov's task taxonomy.
\textit{Estimate Administrative Unit} tasked participants to quantify an administrative unit's associated data value.
We added this task type because it asks directly for the magnitude associated with a region in a single cartogram, whereas all other task types concern the relation between different regions or between different cartograms.
In Table~\ref{tab:tasktypes}, we list all task types used in our experiment, each with a description and an example task.

\definecolor{LightGrey}{rgb}{0.9,0.9,0.9}
\definecolor{MidGrey}{rgb}{0.8,0.8,0.8}

\begin{table*}[!tp]
\renewcommand{\arraystretch}{1.2}
\normalsize
\caption{The seven task types used in our experiment with a sample question for each task type.}
\label{tab:tasktypes}
\centering
\begin{center}
\begin{tabular*}{\textwidth}[h]{ | p{4em} | p{47.7em}|}
\hline
\rowcolor{LightGrey}
\multicolumn{2}{|c|}{Estimate Administrative Unit}\\
\hline
Task & Given a cartogram, participants were required to estimate an administrative unit's associated data value. 
\\ \hline
Example & On the other monitor, you can see a conventional map of Denmark and a population (2018) cartogram.
Estimate the population of Hovedstaden (HS). If you are uncertain, please enter ``NA.''
\\[.4em]
\hline
\rowcolor{LightGrey}
\multicolumn{2}{|c|}{Compare Administrative Units}\\
\hline
Task & Given a cartogram, participants were required to determine whether an administrative unit was larger or smaller than another, and by what magnitude (e.g., population size) they were different.
 \\
\hline
Example & On the other monitor, you can see a conventional map of Sri Lanka and a population cartogram.
Is the population of Hambantota (HB) larger or smaller than the population of Kegalle (KE)? By what magnitude is the population of Hambantota (HB) smaller or larger than Kegalle (KE)? If you are uncertain, please enter ``NA.'' \\[.4em]
\hline
\rowcolor{LightGrey}
\multicolumn{2}{|c|}{Compare Zones}\\
\hline
Task & Given a cartogram divided into two zones (i.e., contiguous sets of administrative units) shown in distinct colors (yellow versus purple), participants were required to determine whether one zone was larger or smaller than the other, and by what magnitude (e.g., population size) they were different. 
\\ \hline
Example & On the other monitor, you can see a conventional map of Belgium and a population cartogram, divided into two zones by color (purple, yellow).
Is the population in the purple region larger or smaller than the population in the yellow region? By what magnitude is the population in the purple region smaller or larger? If you are uncertain, please enter ``NA.''
\\[.4em]
\hline
\rowcolor{LightGrey}
\multicolumn{2}{|c|}{Detect Change in Administrative Unit}\\
\hline
Task & Given two cartograms of the same country, participants were required to estimate the extent to which an administrative unit changed in magnitude (e.g., population size).  
\\ \hline
Example & On the other monitor, you can see a conventional map of India and two cartograms representing the population in 1961 and in 2018.
Is the population of Gujarat (GJ) in 1961 larger or smaller than its population in 2018? By what magnitude is the population of Gujarat (GJ) larger in 1961? If you are uncertain, please enter ``NA.''
\\[.4em] \hline
\rowcolor{LightGrey}
\multicolumn{2}{|c|}{Detect Change in Zone}\\
\hline
Task & Given two cartograms of the same country that was divided into two zones (indicated by the colors yellow versus purple), participants were required to estimate the extent to which a zone changed in magnitude (e.g., population size).
\\ \hline
Example & On the other monitor, you can see a conventional map of New Zealand and two cartograms representing the population in 1991 and 2018.
Is the population of the South Island (yellow) in 1991 larger or smaller than its population in 2018? By what magnitude is the population of the yellow region smaller in 1991? If you are uncertain, please enter ``NA.''
\\[.4em] \hline
\rowcolor{LightGrey}
\multicolumn{2}{|c|}{Cluster}\\
\hline
Task    & Given a cartogram and an administrative unit \textit{U}, participants were required to choose, from a set of four candidates, the administrative unit that had an area most similar to \textit{U}.
\\ \hline
Example & On the other monitor, you can see a conventional map of the United States and a population (2018) cartogram.
Out of the states listed below, which state has a population most similar to Colorado (CO)?       
\\[.4em] \hline
\rowcolor{LightGrey}
\multicolumn{2}{|c|}{Find Top}\\
\hline
Task    & Given a cartogram, participants were required to identify the administrative unit with the largest area from a set of four candidates. 
\\ \hline
Example & On the other monitor, you can see a conventional map of Kazakhstan and a population cartogram.
Which district has the largest population?
\\[.4em] \hline
\end{tabular*}
\end{center}
\end{table*}

\subsubsection{Tasks with Numerical Estimates}
In all tasks requiring numerical estimates (i.e., tasks of the types \textit{Estimate Administrative Unit}, \textit{Compare}, and \textit{Detect Change}), participants had the option of entering ``NA'' (no answer) if they were uncertain.
We chose to include the ``NA'' option because Sischka \textit{et al.}~\cite{sischka_impact_2022} showed that online questionnaires with a forced-answer design (i.e., without the option to indicate a missing answer) lead to higher dropout rates.
Our experiment was supervised in a one-on-one setting, which made dropouts less likely than in Sischka \textit{et al.}’s unsupervised quasi-experiment.
Nevertheless, a forced-answer design would have given participants an incentive to end the experiment quickly by entering arbitrary numbers.
It would have been difficult to detect such answers in the data in hindsight.
Therefore, we expected that the option to choose ``NA'' leads to greater reliability.
We acknowledge that ``NA'' can lead to missing data points because some participants who entered ``NA'' might have been able to estimate the area accurately if they had been forced to enter a number.
However, we mitigate this effect by recruiting a relatively large number of participants (44, see Section~\ref{sec:participants}) and using non-matched and non-paired tests for data analysis (see Section~\ref{sec:data_analysis}).

\subsubsection{Task Presentation}
For each task, a conventional equal-area map was presented alongside the cartogram(s). 
Each administrative unit was filled with the same color on the conventional map and the cartogram(s). 
We used the ColorBrewer palette Dark2 with six colors~\cite{brewer_colorbrewer_2013} and ensured that distinct colors were used for neighboring regions.
On the conventional map, we labeled every administrative unit with two-letter abbreviations. 
The labels simplified the process of locating the administrative units for participants who may have been unfamiliar with the geography of the displayed country.
We also implemented a linked-brushing effect, whereby the color of an administrative unit changed its brightness on both maps when the participant hovered the pointer over the unit on one of the maps~\cite{duncan_task-based_2021}.
The choice of colors and the linked-brushing effect were intended to make it easier for participants to locate an administrative unit on all maps shown during a task~\cite{tingsheng_motivating_2020}.
The screen recordings revealed that all participants deliberately used linked brushing during the cartogram tasks to compare between the labeled administrative units on the equal-area map and the unlabeled administrative units in the corresponding cartogram.

\subsection{Hypotheses}
\label{subsec:experiment_hypotheses}

Prior to the experiment, we hypothesized that legends and grid lines would improve accuracy and increase the task completion rate in all task types with numerical responses (i.e., \textit{Estimate}, \textit{Compare}, and \textit{Detect Change}).
While the total value, printed on every map, enables a rough conversion on a large scale, the legend adds information regarding converting areas to values on a small scale.
We hypothesized that the legend would simplify cartogram reading tasks, even without adding grid lines to the cartogram, because readers could mentally overlay a grid and assess how many legend squares fit into a region.
Nevertheless, we expected that the conversion between legend area and area of a region on the cartogram would require a lower cognitive load when the actual grid lines are visible, leading to more accurate and confident answers.
We also anticipated that additional features might cause slower responses because they encourage participants to make more careful and, thus, time-consuming judgments.
Hence, we made the following hypotheses about numerical-response task types:
\begin{enumerate}[label=\textbf{H\arabic*:}]
\item Additional features would lead to more accurate estimates of magnitudes, conditioned on the task being completed.
\item Additional features would give participants greater confidence in estimating magnitudes and, thus, increase the task completion rate.
\item Participants would need more time in the presence of additional features.
\end{enumerate}
Both task types that did not require numerical responses (i.e., \textit{Find Top} and \textit{Cluster}) could be answered by ranking areas.
While we expected that legends and grid lines would allow participants to make more accurate judgments, we conjectured that these features would not provide direct support for ranking.
\begin{enumerate}[label=\textbf{H\arabic*:}]
\setcounter{enumi}{3}
\item For \textit{Find Top} and \textit{Cluster} tasks, we hypothesized that there would be no significant differences between treatment groups in terms of error rates and response times.
\end{enumerate}
Furthermore, we conjectured that responses to \textit{Estimate Administrative Unit} tasks would not exhibit a systematic trend of overestimating or underestimating areas.
However, we anticipated that legends and grid lines would reduce variability in the responses because these features would provide a visual guide for the estimation.
Therefore, we assumed that the distribution of numerical responses to \textit{Estimate Administrative Unit} tasks would have

\begin{enumerate}[label=\textbf{H\arabic*:}]
\setcounter{enumi}{4}
\item a mean near the true magnitude of the administrative unit to be estimated for all treatments.
\item less variability in treatments with additional features (i.e., static legend, grid lines, and selectable legend) than in the no-features treatment.
\end{enumerate}

\subsection{Data Sets}
\label{subsec:datasets}

We generated cartograms of 28 different randomly chosen countries.
These countries are listed in Section~1 of the supplemental text, available online.
All maps showed boundaries of first-level administrative units, which is typical of cartograms that users encounter in real life.
The mean number of administrative units was 18.4 and the median 16.0.
We acknowledge that the number of administrative units varies (standard deviation 10.4).
However, the variability is consistent with realistic use cases of cartograms.
Participants encountered each cartogram only once during the experiment because we wanted to mitigate any learning effects.

To generate the cartograms, we used several types of data, such as population, GDP, and COVID-19 cases. 
For the \textit{Detect Change in Administrative Unit} and \textit{Detect Change in Zone} questions, where the task was to compare two cartograms, we used data sets of the same type of data for two different years (e.g., population in 1985 and population in 2018) and showed the participants both cartograms next to each other.
When developing questions for tasks that involved comparing administrative units
(i.e., \textit{Compare Administrative Units} and \textit{Cluster}),
we chose administrative units with similar areas to make these questions challenging and comparable in difficulty.
For cartograms
that
were divided into two zones for the \textit{Detect Change in Zones} and \textit{Compare Zones} tasks,
we attempted to divide the cartograms into equal halves to provide participants also with comparably challenging tasks.
All cartograms were generated using the web application \href{https://go-cart.io}{go-cart.io}, which uses the fast flow-based algorithm proposed by Gastner~\textit{et al.}~\cite{gastner_fast_2018}.

\subsection{Participants}
\label{sec:participants}
We recruited 44 participants, all students of Yale-NUS College.
Approval for research on human subjects was obtained from the Yale-NUS Undergraduate Research Ethics Committee (case ID RI-00000620).
Out of the total sample, 20 participants were female, 22 were male, and 2 preferred not to answer. 
The mean age of the participants was 20.0, and all participants were between 18 and 24 years of age. 
The experiment required participants to distinguish between map regions filled with different colors; thus, we used the Ishihara color perception test to ensure that the participants were not color blind.
Because all participants were able to identify the correct numbers on the Ishihara test plates, we believe that none of the participants was color blind.
Consequently, the results of all participants were included in our analysis.

Before participants started working on any cartogram task, we asked them to assess their familiarity with maps and cartograms on a 5-point Likert scale (1=``Not familiar at all'' and 5=``Extremely familiar'').
The mean rating was 3.0 for familiarity with maps in general and 1.9 for cartograms.
The low familiarity with cartograms is likely to be representative of most casual cartogram users.

We acknowledge that a limitation of our experiment is that all participants were college students. 
Therefore, our results may only be representative of a younger, more educated population. 
Previous cartogram evaluation studies also faced this limitation \cite{dent_communication_1975, nusrat_evaluating_2016, rittschof_learning_1996}. 
However, the tasks in our experiment did not require specialized knowledge and can be grasped by teenagers and adults with normal vision (possibly with correction). 
Thus, we believe that our results can be generalized to a larger population.  

\begin{figure*}[!t]
\centering
\includegraphics[width=6in]{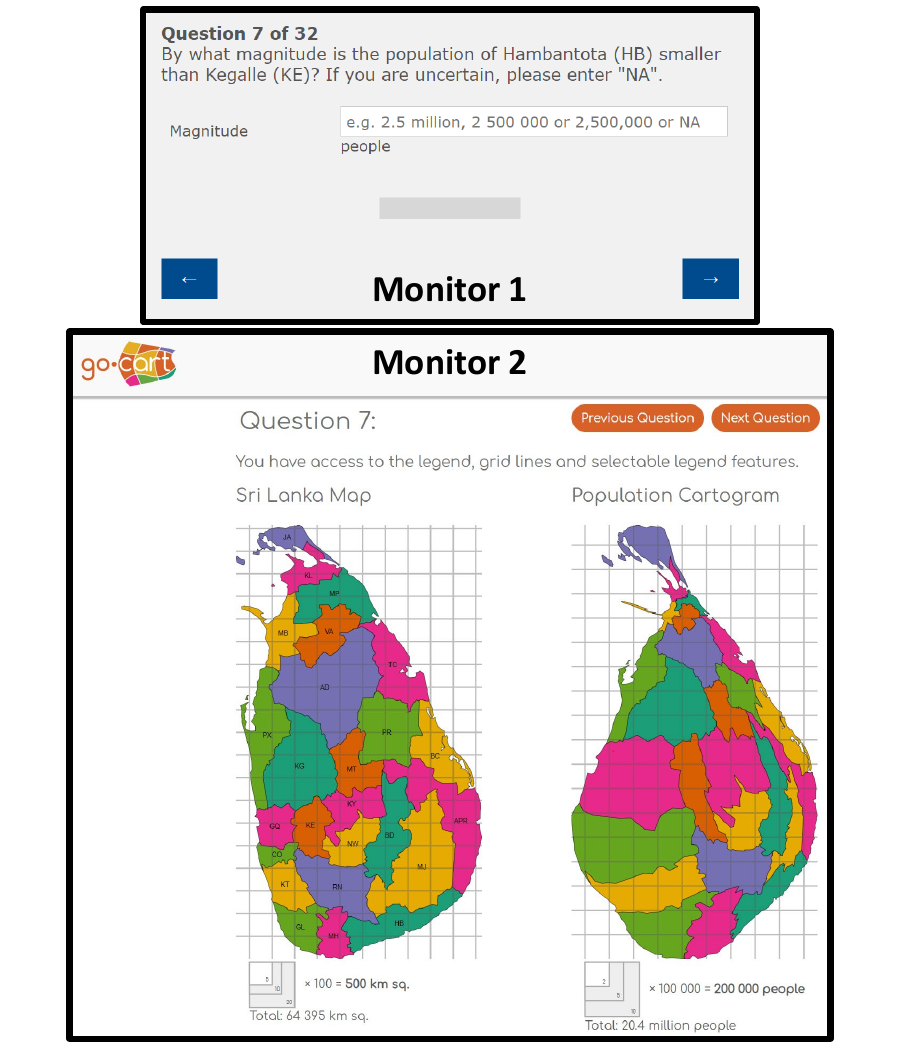}
\caption{Experimental setup with two monitors.
In the experiment, monitor 1 was placed immediately to the right of monitor 2.
On monitor 1, participants read the current question and entered their response.
Monitor 2 showed the graphical user interface that contains a conventional map and one or two cartograms of the same country.} 
\label{fig:experiment_setup}
\end{figure*}

\subsection{Procedure and Design}
\label{sec:procedure_and_design}
The experiment consisted of four parts.
During all four parts, there was always an experiment supervisor present to oversee the procedure in a one-on-one setting.

\begin{figure*}[!ht]
  \centering
    \includegraphics[width=6.5in]{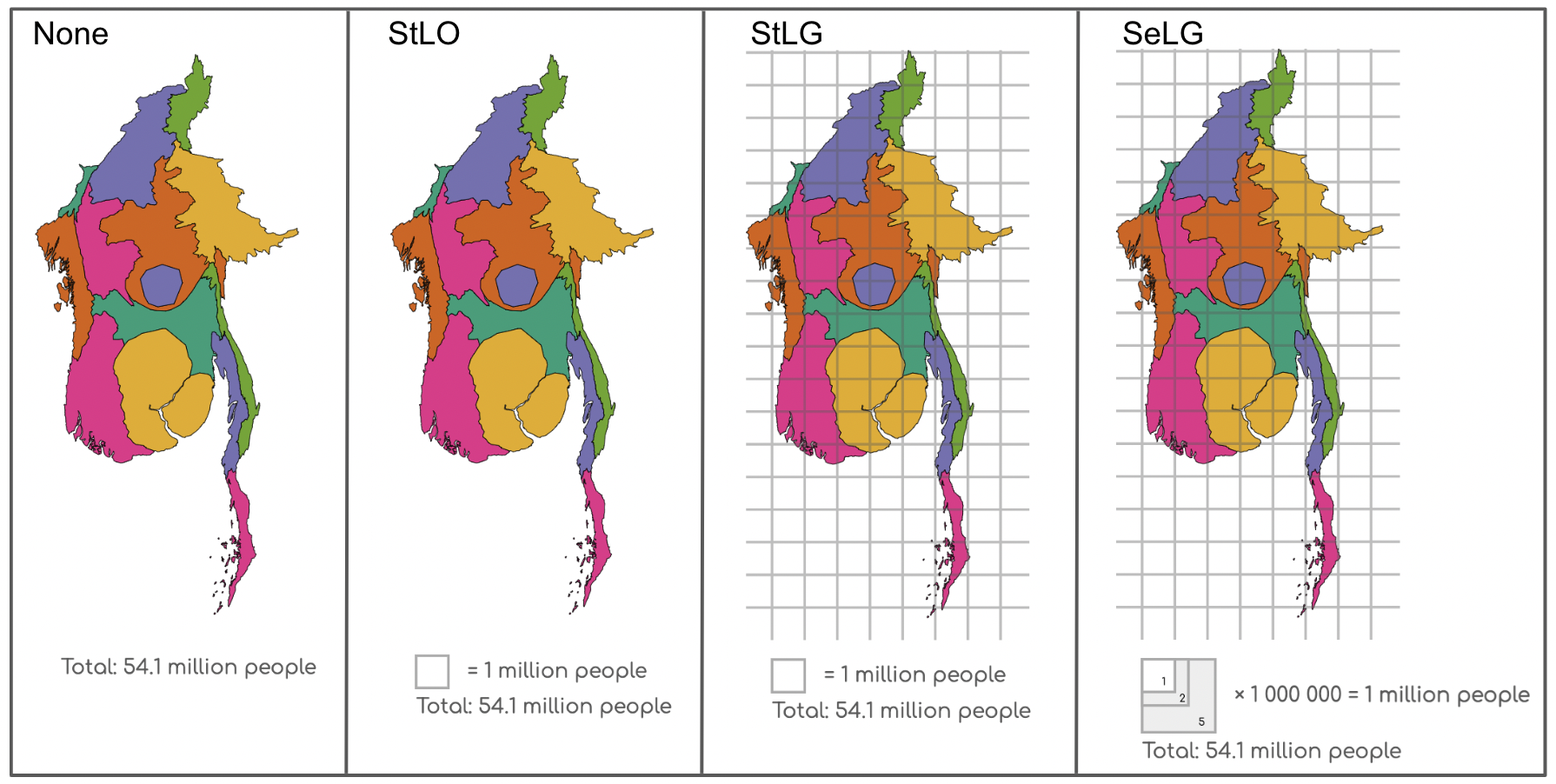}
  \caption{%
  Illustration of the four treatments investigated in this study, using Myanmar's population cartogram as an example.
  (1) None: Neither legend nor grid lines.
  (2) StLO: Static legend only.
  (3) StLG: Static legend and grid lines.
  (4) SeLG: Selectable legend and grid lines.}
  \label{fig:four_treatments}
\end{figure*}

\begin{enumerate}[label={(\arabic*)}]
\item
\label{experiment:intro}
\textbf{Introduction:} At the start of the experiment, participants read an information sheet and then signed a form to indicate that they consented to participating in the experiment.

Next, participants sat in front of two liquid-crystal display monitors, each with a resolution of $1920 \times 1080$, and watched a five-minute video that introduced them to cartograms and provided details about the experimental procedure. 
During the video, participants could pause, rewind, and ask for clarification as they wished.
After the video, the experiment supervisor set up the two displays as follows (Fig.~\ref{fig:experiment_setup}).
Monitor 1 displayed a Qualtrics XM survey, where participants read task descriptions and entered their answers. 
Monitor 2, positioned immediately to the left of monitor 1, displayed a web-based graphical user interface that showed the conventional maps and cartograms.
We used separate monitors for questions and maps so that participants could see cartograms and legends at a large (i.e., full-screen) scale.
If we had used only one screen, either the cartograms would have been too small to interact comfortably with the selectable legend, or participants would have had to spend time on toggling between questions and maps, unrelated to the cartogram task.
Both screens were placed in immediate proximity at eye level to minimize the distance needed for eye movements.
  
With this setup, participants answered a practice question, which consisted of two tasks:
\begin{itemize}
\item
  Participants had to test the linked-brushing feature (i.e., parallel highlighting of corresponding regions on an equal-area map and a cartogram).
  Then they had to confirm that they were comfortable using this feature before they were allowed to proceed to the task-based questions.
\item
  Participants had to interpret the highlighted legend square and enter the value associated with the largest square.
  The value had to be correct before participants could advance to the task-based questions.
\end{itemize}
These two tasks covered all interactive map features available to participants during the experiment.
Because no participant needed help from the supervisor to complete the practice question, it is unlikely that participants' response times during the task-based questions were affected by learning how to use the interactive features.
The cartogram and country used in this practice question were not reused in later questions.

\item
\label{experiment:prelim}
\textbf{Preliminary questions:} Participants answered questions about their age, gender, and level of education.
Participants also rated their familiarity with maps and cartograms.
Thereafter, we conducted a color perception test with four Ishihara test plates. 

\item
\label{experiment:tasks}
\textbf{Cartogram tasks:} Participants answered 28 task-based questions.
Each question required them to read a conventional map and one or two cartograms.
They were provided with scratch paper and a pocket calculator.
We also informed participants that they could take as much time as they needed for each question and that the time they took to complete each question would be recorded.

We used a 7 $\times$ 4 within-subject experiment design.
There were 7 task types, as listed in Table \ref{tab:tasktypes}.
Each task type was combined with one of the following 4 treatments, characterized by the features displayed in the cartogram(s) for the task:

\begin{itemize}
\item Neither legend nor grid lines
\item Static legend only
\item Static legend with grid lines
\item Selectable legend with grid lines
\end{itemize}

Figure \ref{fig:four_treatments} illustrates these 4 treatments using Myanmar's population cartogram.
Each participant encountered each combination of task type and treatment only once.
We randomly divided the 44 participants into 4 groups, and the order in which the tasks appeared was the same for all groups.
However, participants in different groups encountered the 4 experimental conditions at different times, as per the Latin square design~\cite{grant_latin_1948}, which is a standard experimental design to counterbalance order effects (e.g., learning or fatigue)~\cite{corriero_counterbalancing_2017}.
Every question used a unique map and the 4 groups together covered all treatments for this map.
Section 1 of the supplemental text, available online, lists the order in which combinations of countries, task types, and features appeared during the experiment.

\item
\label{experiment:attitude}
\textbf{Attitude study:} We posed four free-response questions in which participants were asked to briefly describe the different strategies they used to perform the tasks in each of the four treatments (i.e., no features, static legend only, static legend with grid lines, and selectable legend with grid lines).
The free-response questionnaire was conducted immediately after the cartogram tasks to ensure that participants had fresh memory of their strategies.
Participants had to write their descriptions in text boxes displayed on monitor 1, which previously had shown all survey questions.

Next, participants rated the aesthetics and effectiveness of static legends, static grid lines, and selectable legends with grid lines in a semantic differential test that we adapted from the tests used by Dent~\cite{dent_communication_1975} and Nusrat~\textit{et al.}~\cite{nusrat_evaluating_2016}.
We devised seven pairs of words and phrases with opposite meaning:
\begin{itemize}
\item Hindering \textendash~Helpful
\item Redundant \textendash~Essential
\item Difficult to understand \textendash~Easy to understand
\item Showing magnitude poorly \textendash~Showing magnitude clearly
\item Does not form an immediate impression \textendash~Forms an immediate impression
\item Ugly \textendash~Elegant
\item Conventional \textendash~Innovative
\end{itemize}
For each of the three features and each of the seven phrase pairs, participants indicated their attitude on a 5-point Likert scale.
\end{enumerate}

The median duration of the experiment was 44 minutes (interquartile range: 15 minutes) with one high outlier at 82 minutes.
The instructional video used in part~\ref{experiment:intro} and the complete list of questions used in parts~\ref{experiment:prelim}--\ref{experiment:attitude} are available as supplemental material for this article from the journal's website.

\subsection{Data Analysis}
\label{sec:data_analysis}

Responses to the seven task types in the experiment belonged to one of two categories: numerical responses or the name of an administrative unit.
Five of the seven task types belong to the former category (\textit{Estimate Administrative Unit}, \textit{Compare Administrative Units}, \textit{Compare Zones}, \textit{Detect Change in Administrative Unit}, and \textit{Detect Change in Zone}), while two belong to the latter (\textit{Find Top} and \textit{Cluster}).
For the five task types that required numerical responses, we wanted to compare the participants' task completion rates, the accuracy of their responses, and response times when performing the tasks under the four treatments listed in Section \ref{sec:procedure_and_design}.
For the two task types that did not require numerical responses, we wanted to compare error rates and response times under the four treatments.

The five task types with numerical responses can be divided into two subcategories.
In the first subcategory (\textit{Estimate Administrative Unit}), the correct answer had to be a positive number.
However, in the second subcategory (comprising both \textit{Compare} and both \textit{Detect Change} task types), the correct answer could be positive or negative.
Therefore, responses to the second subcategory were in two parts.
First, participants had to answer whether the focal region had a larger or smaller area than the reference region (i.e., they had to determine the sign of the difference in area).
Subsequently, participants entered the magnitude of the difference.

To compare the participants' accuracy for the five task types that required numerical responses, we first normalized the numerical data as follows: 

\begin{enumerate}
\item If a participant answered the first part of a two-part question (e.g., \textit{Compare Administrative Units}) incorrectly, we inverted the sign of the numerical response given in the second part.
For example, suppose that a participant thought that region $A$ is larger than region $B$ by $5000$, but the correct answer is that region $A$ is smaller than region $B$ by $3000$.
The two parts of the participant's answer would have been recorded as (i) ``Larger,'' and (ii) ``$5000$.''
Because (i) was incorrect, we inverted the sign of (ii) to obtain $-5000$.
Thus, the participant would have been off by $-5000 \mbox{ (response)} - 3000 \mbox{ (correct answer)} = -8000$.
This step allows us to properly account for the difference between the participant's numerical response and the correct answer.
A positive sign of the sign-adjusted response indicates that the direction of the estimate is correct.

\item We then normalized the response using the below formula: 
\[ \mbox{normalized response} = \frac{\mbox{response} - \mbox{correct answer}}{\mbox{correct answer}}\ . \]
Normalized responses for \textit{Estimate Administrative Unit} tasks had to be greater than or equal to $-1$.
For all other task types, values less than $-1$ were possible.
\end{enumerate}

After normalizing the numerical responses, we could meaningfully aggregate the results for different tasks of the same task type, even if the magnitude of the answer differed between the tasks.
However, even after the above two steps, many of the distributions of the normalized responses were skewed.
Therefore, we used the Kruskal--Wallis test for differences between treatments.
The Kruskal--Wallis test is a non-parametric test that does not assume that the data are normally distributed, matched, or paired~\cite{mcknight_kruskal-wallis_2010}.
The test statistic is $\chi^2$-distributed with 3 degrees of freedom.
For post-hoc analysis, we used pairwise Mann--Whitney U tests with Bonferroni--Holm correction~\cite{abdi_holms_2010, mcknight_mann-whitney_2010}.

In the five task types with numerical responses, participants had the option to enter ``NA'' (no answer) instead of a number if they were uncertain.
We define the task completion rate as percentage of non-``NA'' responses.
To assess whether the participants' task completion rate differed between treatments, we used Cochran's Q test~\cite{mccrum-gardner_which_2008}.
In the post-hoc analysis, we used pairwise McNemar tests with Bonferroni--Holm correction~\cite{abdi_holms_2010, lachenbruch_mcnemar_2005}.
For significant pairwise differences, we used the odds ratio to measure the effect size and calculated confidence intervals using the method proposed by Fay~\cite{fay_two-sided_2010}.

To calculate the distribution of response times for the five numerical-response task types, we excluded the response times of participants who did not complete the task (i.e., entered ``NA'' for the area estimate).
Because the distributions of response times were right-skewed, we used the Kruskal--Wallis test to assess whether there are differences between treatments.
If the Kruskal--Wallis test rejected the null hypothesis that there is no difference, we used pairwise Mann--Whitney U tests with Bonferroni--Holm correction for post-hoc analysis.

For the task types that did not require a numerical response (i.e., \textit{Find Top} and \textit{Cluster}), we treated the response as binary data (correct versus incorrect) and applied Cochran's Q test, used pairwise McNemar tests for post-hoc analysis, and determined the odds ratio and confidence intervals. 
To compare response times for these two task types, we excluded the response times of participants who answered the question incorrectly.
Similar to comparing response times for numerical-response task types, we used the Kruskal--Wallis test.
For post-hoc analysis, we used pairwise Mann--Whitney U tests with Bonferroni--Holm correction.

For \textit{Estimate Administrative Unit} tasks, we performed additional data analysis.
We were interested in whether there was bias in overestimating or underestimating the magnitude.
The distribution of the residuals (i.e., the normalized responses minus the mean conditioned on the treatment) failed the Shapiro--Wilk test for normality ($W = 0.96$, $p < 10^{-3}$).
Thus, we applied the non-parametric Wilcoxon signed-rank test for a difference between zero and the pseudomedian of the normalized responses with Bonferroni--Holm correction.
We also applied the Fligner--Killeen test to determine whether the variability of the numerical responses differed among the four treatments.
For post-hoc analysis, we applied Ansari--Bradley tests for differences in scale parameters with Bonferroni--Holm correction.

In all the tests, we considered $p$-values $\leq 0.05$ as statistically significant.
As pointed out by Cumming~\cite{cumming_understanding_2011}, confidence intervals communicate the range of 
uncertainty more clearly than $p$-values alone.
Therefore, we complement the $p$-values with confidence intervals for all pairwise comparisons in the supplemental text, available online.
The data and R scripts used for our statistical analysis are publicly available at \url{https://github.com/kvelon/cartogram-legend-effectiveness}.

\begin{figure*}[tp]
\centering
\includegraphics[width=0.75\textwidth]{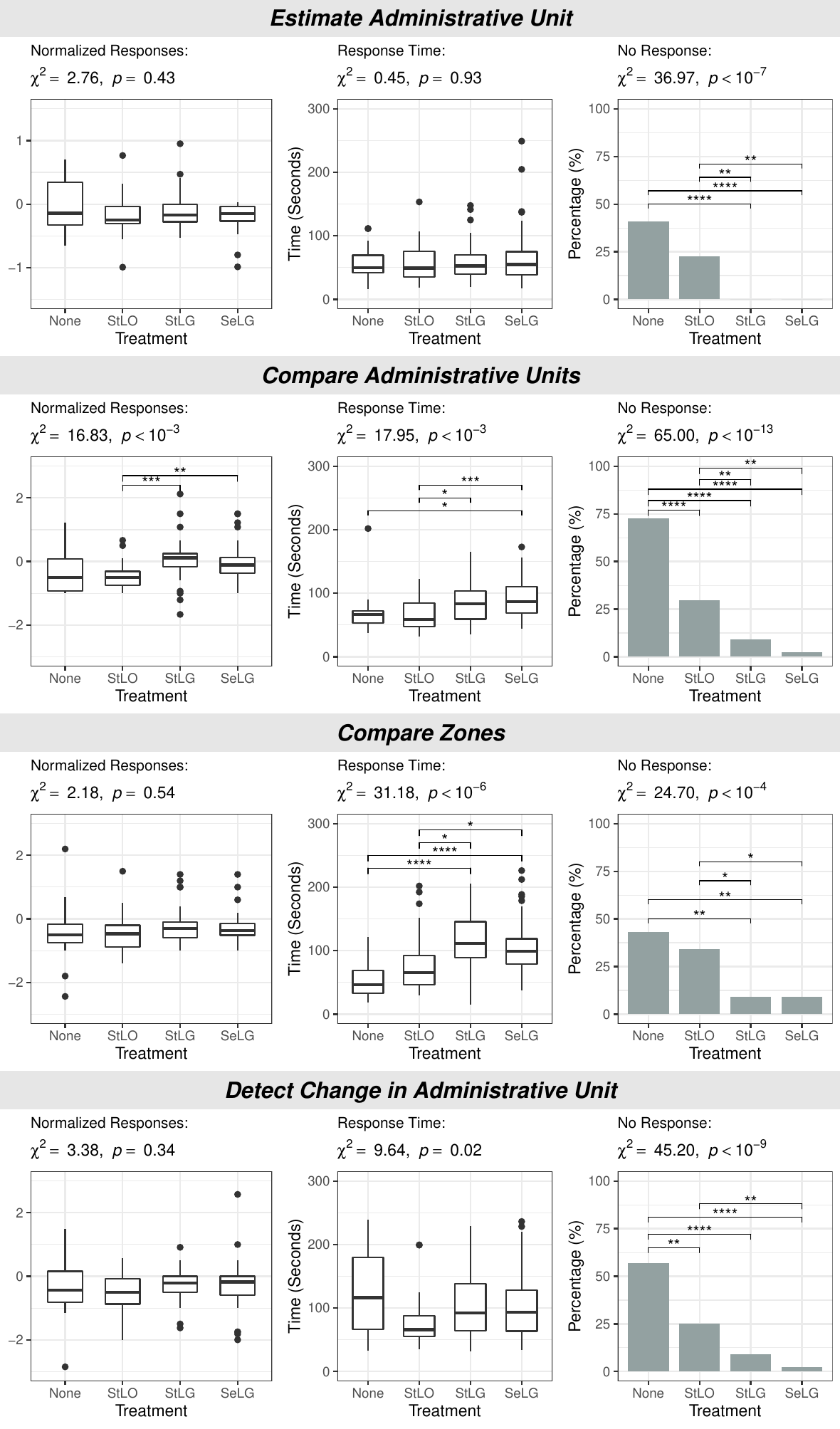}
\end{figure*}

\begin{figure*}[tp]
\centering
\includegraphics[width=0.75\textwidth]{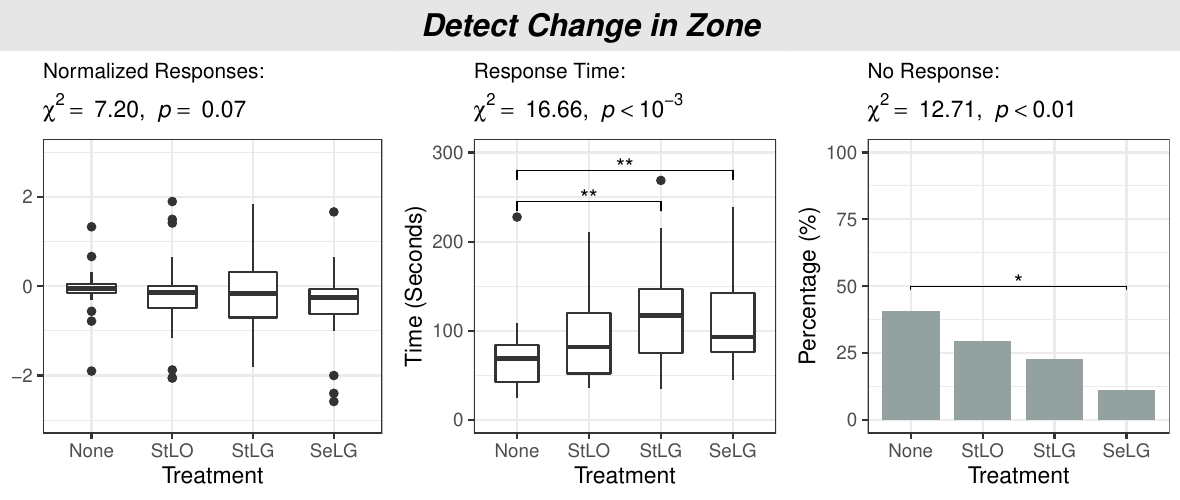}
\caption{Distributions of the accuracy of responses, response times, and missing responses for the five task types that required numerical responses.
The following abbreviations are used in the axis labels.
StLO: Static legend only.
StLG: Static legend with grid lines.
SeLG: Selectable legend with grid lines.
Horizontal brackets in the plots indicate significant differences between pairs of treatments at a significance level of 0.05.
Asterisks above the brackets indicate the $p$-value.
$\ast$: $p$-value $\leq$ 0.05.
$\ast\ast$: $p$-value $\leq$ 0.01.
$\ast\ast\ast$: $p$-value $\leq$ 0.001.
$\ast\ast\ast\ \ast$: $p$-value $\leq$ 0.0001.}
\label{fig:quant_tasks}
\end{figure*}

\section{Results}

\subsection{Accuracy of Responses, Response Times, and Task Completion Rates}

\subsubsection{Task Types with Numerical Responses}
The results of our data analysis for the five task types with numerical responses are shown in Fig.~\ref{fig:quant_tasks}.
Tabular summaries of the results can be found in the supplemental text, available online.
In the supplement, we also report confidence intervals of significant pairwise differences.
The data analysis, outlined in Section~\ref{sec:data_analysis}, led to the following findings.

\begin{itemize}
\item \textbf{Estimate Administrative Unit}:
Differences between treatments had no significant effect on the average accuracy.
However, the Fligner--Killeen test rejected the null hypothesis of equal standard deviation in the normalized responses [$\chi^2(3) = 12.21$, $p < 0.01$].
Pairwise Ansari--Bradley tests revealed that the selectable-legend-with-grid-lines treatment had a significantly smaller standard deviation ($0.21$) than the no-features treatment ($0.38$).
Wilcoxon signed-rank tests detected that the median was different from zero in all but the no-features treatment.
Participants generally tended to underestimate the areas.
The median ranged from $-0.14$ for the no-features treatment to $-0.25$ for the static-legend-only treatment.

There was no evidence of a dependence of the response time on the treatment, but there was a significant effect on the task completion rate [$\chi^{2}(3) = 36.97, p < 10^{-7}$]. 
When participants had access to a static legend with grid lines or a selectable legend with grid lines, all participants completed the task. 
The post-hoc analysis of the participants' task completion rate confirmed that there were significant differences between the treatments without grid lines (lower completion rate) and those with grid lines (higher completion rate).

\item \textbf{Compare Administrative Units}: For this task type, differences between treatments had a significant effect on the accuracy of the responses [$\chi^{2}(3) = 17.63, p < 10^{-3}$], response time [$\chi^{2}(3) = 17.10, p < 10^{-3}$], and the task completion rate [$\chi^{2}(3) = 65.00, p < 10^{-13}$]. 
Post-hoc analyses for the accuracy and response time showed pairwise differences between the static-legend-only treatment and the two treatments with grid lines.
Hence, grid lines made responses more accurate but slower.
For the task completion rate, we found pairwise differences between the no-features treatment and the other three treatments, as well as between the static-legend-only treatment and the two treatments with grid lines, which made participants more likely to complete the tasks.
Furthermore, we found that the majority of participants did not complete the task when they did not have any features (72.7$\%$).

\item \textbf{Compare Zones}: We did not observe a significant effect of the treatments on the accuracy of the response.
However, we did observe a significant effect on the response time [$\chi^{2}(3) = 31.18, p < 10^{-6}$] and the task completion rate [$\chi^{2}(3) = 24.70, p < 10^{-4}$]. 
For the response time and task completion rate, we observed pairwise differences between the no-features treatment (fastest responses and lowest completion rate) and the two treatments with grid lines.
We also observed a difference between the static-legend-only treatment and the two treatments with grid lines (slowest responses and highest completion rate).

\item \textbf{Detect Change in Administrative Unit}: Differences between treatments had no significant effect on accuracy, but we observed a significant effect on the response time [$\chi^{2}(3) = 9.64, p = 0.02$] and task completion rate [$\chi^{2}(3) = 45.20, p < 10^{-9}$]. 
A majority of participants (56.8\%) did not complete the task when they did not have any additional features.
Post-hoc analysis revealed pairwise differences between the no-features treatment and the other three treatments, and between the static-legend-only treatment and the selectable-legend-with-grid-lines treatment, which made participants more likely to complete the task.

\item \textbf{Detect Change in Zone}: We did not observe a significant effect of the different treatments on the accuracy of the response [$\chi^{2}(3) = 7.20, p = 0.07$].
However, we found a significant effect on the response time [$\chi^{2}(3) = 16.66, p < 10^{-3}$] and task completion rate [$\chi^{2}(3) = 12.71, p < 0.01$]. 
For the response time, post-hoc analysis revealed pairwise differences between the no-features treatment (fastest responses) and the two treatments with grid lines (slowest responses).
For the task completion rate, we found a pairwise difference between the no-features treatment (lowest completion rate) and the selectable-legend-with-grid-lines treatment (highest completion rate).
\end{itemize}

\subsubsection{Task Types without Numerical Responses}

\begin{figure*}[ht]
  \centering
    \includegraphics[width=\textwidth]{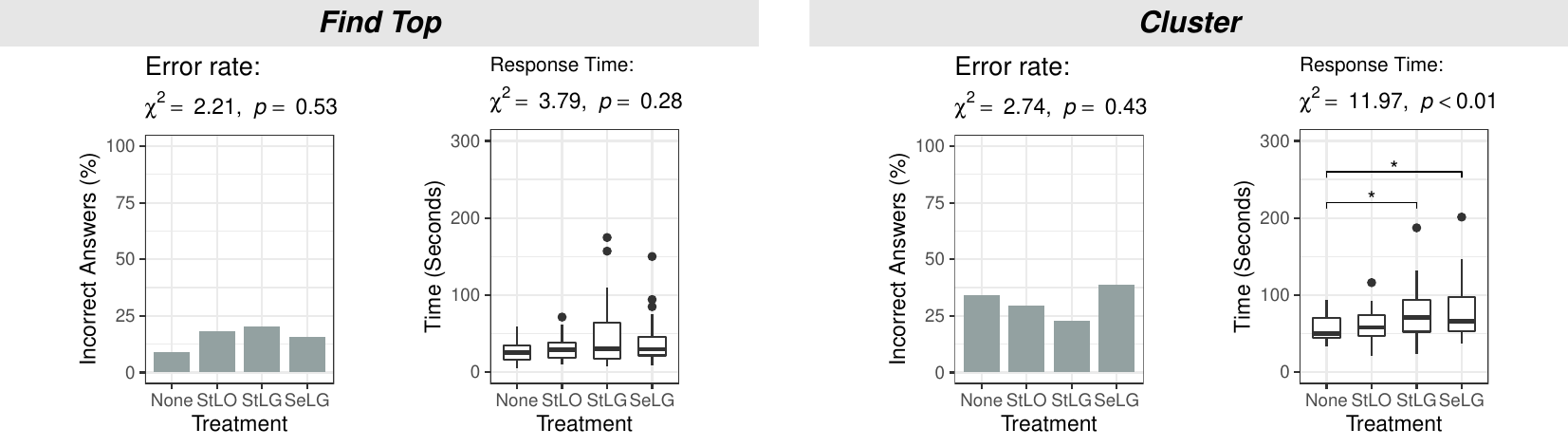}
  \caption{Distributions of error rates and response times for the two task types that did not require numerical responses.
  The following abbreviations are used in the axis labels.
  StLO: Static legend only.
  StLG: Static legend with grid lines.
  SeLG: Selectable legend with grid lines.
  Horizontal brackets in the plots indicate significant differences between pairs of treatments at a significance level of 0.05.
  An asterisk above a bracket indicates a $p$-value in the interval [0.01, 0.05].}
  \label{fig:non_quant_tasks}
\end{figure*}

For non-numerical task types, we measured the error rates and response times (Fig.~\ref{fig:non_quant_tasks}).
We did not give participants the option to skip non-numerical tasks; the tasks had to be completed to move on to the next question.

\begin{itemize}
\item \textbf{Find Top}: Differences between treatments had no significant effect on the error rate or on the response time for tasks of this type.

\item \textbf{Cluster}: We did not observe a significant effect of the treatments on the error rate.
However, we did observe a significant effect on the response time [$\chi^{2}(3) = 11.97, p < 0.01$].
Post-hoc analysis of the response time revealed pairwise differences between the no-features treatment (fastest responses) and both treatments with grid lines (slowest responses).
\end{itemize}

\subsection{Hypotheses}
Regarding the hypotheses made prior to the experiment (see Section~\ref{subsec:experiment_hypotheses}), we draw the following conclusions:

\begin{itemize}
\item \textbf{H1 is rejected} because the treatment groups had no significant effect on the participants' accuracy in four out of five numerical-response task types.
\item \textbf{H2 and H3 are supported}.
For task types that required numerical responses, legends and grid lines tended to increase the task completion rates and response times.
\item \textbf{H4 is partially supported}. As expected, we did not observe any significant effect of the treatments on the error rates of the \textit{Find Top} and \textit{Cluster} tasks.
However, we found evidence that additional features slowed down the responses to \textit{Cluster} tasks.
\item \textbf{H5 is rejected} because our experiment revealed a tendency to underestimate the true magnitude in \textit{Estimate Administrative Unit} tasks when a legend was available.
\item \textbf{H6 is partially supported} by our results because we observed that the standard deviation of normalized responses decreased when a selectable legend with grid lines was available in \textit{Estimate Administrative Unit} tasks.
\end{itemize}

\begin{figure}
\centering
\includegraphics[width=0.5\textwidth]{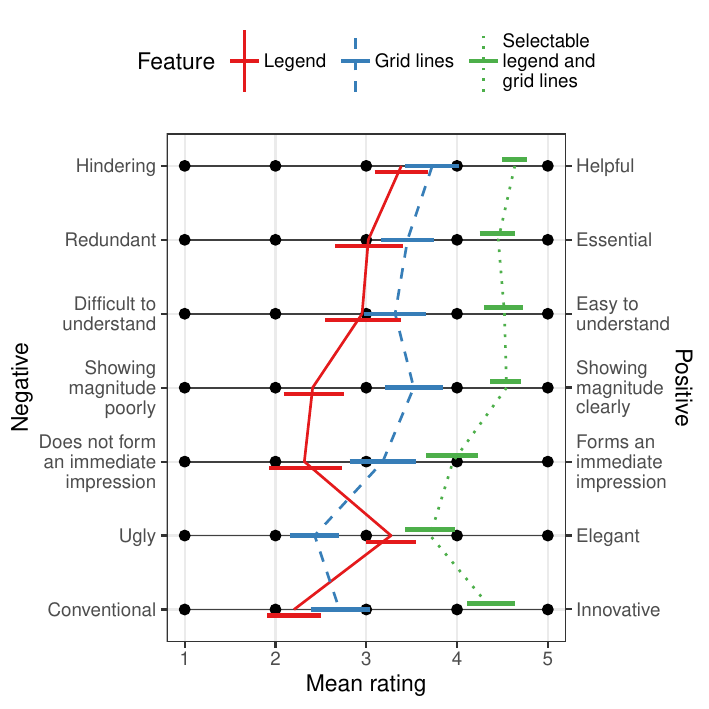}
\caption{Mean ratings in the attitude study conducted after the task-based questions.
Horizontal bars are bootstrap estimates of the 95\% confidence intervals.}
\label{fig:attitude_results}
\end{figure}

\subsection{User Preference and Subjective Rating of Features}
\label{sec:user_preferences}

In the last part of the experiment, we asked the participants to rate the aesthetics and effectiveness of the three features (i.e., legend, grid lines, and selectable legend with grid lines) on a 5-point Likert scale using the phrase pairs listed at the end of Section~\ref{sec:procedure_and_design}.
In Fig.~\ref{fig:attitude_results}, we show the mean ratings for each phrase pair and for each feature. 
The legend feature received the lowest mean rating among the three features for six out of seven phrase pairs (mean of all seven phrase pairs: 2.80).
The selectable legend with grid lines had the highest mean rating (mean of all seven phrase pairs: 4.31). 

In free-response style questions at the end of the experiment, participants described their strategies under different treatment conditions.
For the no-features treatment, all 44 participants entered an answer that can be construed as a strategy in a broad sense.
21 participants wrote that they had first estimated the proportion of the total area that had been occupied by the administrative unit or zone mentioned in the question.
They then multiplied this proportion with the total value of the mapping variable stated below the legend.
However, 3 participants wrote that they were unable to perform the tasks without access to the features; 4 participants described their approach as ``guessing,'' and 4 more participants described the task as ``difficult'' or ``next to impossible.''

For the static-legend-only treatment, 43 out of 44 participants described a strategy.
Two different strategies were indicated most frequently.
The first strategy, described by 16 participants, estimated the number of squares that could fit within an area; these participants then multiplied this number with the legend value. 
The second strategy, chosen by 11 participants, was to disregard the legend and apply the same method many participants used when they had no features; that is, they estimated the proportion of the total area and then multiplied the proportion with the total value of the mapping variable.
For the static-legend-with-grid-lines treatment, one participant did not describe a strategy.
38 out of the remaining 43 participants responded that they had performed the task by using the grid lines to count the number of squares covering the regions mentioned in the question.

Finally, for the selectable-legend-with-grid-lines treatment, 3 participants did not describe a strategy.
Out of the remaining 41 participants, 15 stated that they had chosen the legend size that had the closest fit for the region mentioned in the task; they then counted the number of squares.
The ``best'' fit was a judgment that differed from individual to individual and from question to question. 
7 participants generally chose the largest of the three legend sizes to minimize the number of squares that needed to be counted. 
Another strategy, which was adopted by 3 participants, was to choose a size that would minimize the number of partially filled squares.

\section{Discussion}

\subsection{Selectable Legends with Grid Lines Do Not Make Estimates More Accurate but More Consistent}

The median of the normalized responses to the \textit{Estimate Administrative Unit} tasks was significantly smaller than zero when a legend was available.
The tendency to underestimate the ratio of a larger area (an administrative unit) to a smaller area (a legend symbol) is consistent with results from psychophysics about area perception~\cite{flannery_relative_1971}.
Participants tended to underestimate the magnitude even in the presence of grid lines.
A plausible explanation is that participants might have judged the area based on a count of squares that were completely contained in the administrative unit, and squares that were only partly inside the administrative unit may have been omitted.

While the median response to \textit{Estimate Administrative Unit} tasks did not become more accurate, the variability became smaller when participants had access to a selectable legend with grid lines.
In our opinion, the smaller variability and, hence, greater consistency of the estimates more than offset the negative consequences of moderately underestimating the displayed magnitudes.
An interesting question for future research is whether the tendency toward underestimation could be corrected using the ``apparent magnitude scaling'' technique~\cite{flannery_relative_1971}.

\subsection{Effect of Having a Static Legend without Grid Lines}

Static legends without grid lines did not seem to have a statistically significant impact on the accuracy of the participants' responses when compared to the no-features treatment.
Moreover, the legend did not significantly affect participants' response times compared to the no-features treatment. 
However, we observed that the legend-only treatment allowed participants to be significantly faster than the two treatments with grid lines for the \textit{Compare Administrative Units} and \textit{Compare Zones} tasks. 
Faster responses are likely a consequence of the participants' strategies for performing the tasks when only a legend was available. 
The first strategy involved estimating how many squares were needed to cover an administrative unit or zone; the second strategy disregarded the legend altogether and simply involved estimating the proportion of the total area occupied by a region.
These two strategies did not involve meticulous counting of squares.
Consequently, the static-legend-only treatment was faster than the two treatments with grid lines.

Compared to the no-features treatment, the legend significantly reduced the number of participants who did not complete \textit{Compare Administrative Units} and \textit{Detect Change in Administrative Unit} tasks. 
Notably, the legend made it statistically significantly more likely to complete tasks of these two administrative-unit-based task types but not their sibling task types for zones (i.e., \textit{Compare Zones} and \textit{Detect Change in Zone}). 
A plausible explanation is that the administrative-unit-based task types were more difficult when using the strategy of estimating proportions because administrative units typically made up a much smaller proportion of the total area than zones.

In the attitude study, participants considered the legend feature to be more elegant than grid lines (mean rating of 3.27 versus 2.43), presumably because the legend was less obtrusive.
However, participants gave lower ratings to the legend than to the grid lines for all other phrase pairs.
A plausible explanation for the moderately negative attitude toward the legend is that it appeared some distance below the cartogram, whereas grid lines were directly superimposed on the cartogram.

\subsection{Effect of Having a Static Legend with Grid Lines}

Compared to the no-features treatment, the combination of a static legend with grid lines was not effective in improving the participants' accuracy.
The only statistically significant effect observed was that this treatment allowed participants to give more accurate responses compared to the legend-only treatment for \textit{Compare Administrative Units} tasks. 
The box plot of the normalized responses for \textit{Compare Administrative Units} in Fig.~\ref{fig:quant_tasks} shows many outliers in the static-legend-with-grid-lines treatment.
Therefore, the positive effect of the static legend with grid lines is counterbalanced by more outliers in both directions.

This treatment also caused participants to be slower while performing \textit{Detect Change in Zone} and \textit{Cluster} tasks compared to the no-features treatment, as well as slower in performing \textit{Compare Zones} tasks compared to the no-features treatment and legend-only treatment. 
As mentioned in Section~\ref{sec:user_preferences}, when participants had no additional features or only a legend for \textit{Compare Zones} or \textit{Detect Change in Zone} tasks, many used the strategy of estimating the proportion of the total area occupied by the zone. 
However, with grid lines, most participants used the strategy of counting the number of squares that covered the zone.
The second strategy is more time-consuming, and we believe that it explains why participants were slower with this treatment for these two task types.

The static legend with grid lines had a positive effect on participants' task completion rate.
In four out of the five task types where participants were required to provide numerical responses (\textit{Estimate Administrative Unit}, both \textit{Compare} task types, and \textit{Detect Change in Administrative Unit}), we observed that this treatment significantly increased the task completion rate compared to the no-features treatment.
In fact, all participants completed the \textit{Estimate Administrative Unit} tasks.

\subsection{Effect of Having a Selectable Legend with Grid Lines}

The selectable legend with grid lines did not appear to have an effect on accuracy when compared to the no-features treatment. 
However, we did observe a statistically significant effect on accuracy compared to the static-legend-only treatment in the \textit{Compare Administrative Units} tasks. 
The likely reason is that, with three legend sizes, participants were able to choose a legend size that is closest to the sizes of the administrative units that needed to be compared, thereby making it easier to spot differences in their areas.
Compared to having no features, participants were significantly slower in performing \textit{Compare Zones}, \textit{Detect Change in Zone}, and \textit{Cluster} tasks when given access to a selectable legend with grid lines.
We believe that participants were slower because, with access to a selectable legend, they tended to select a legend size and then counted the number of squares that fit into an area.
This strategy is more time-consuming than the strategy of roughly estimating proportions, which did not utilize any of the features. 

The selectable legend with grid lines made participants more likely to complete the task compared to the treatment without additional features. 
We observed a statistically significant effect in all five task types that required numerical responses. 
Additionally, in these five task types, no treatment had a percentage of missing responses lower than the selectable-legend-with-grid-lines treatment.
Compared to the legend-only treatment, the selectable legend with grid lines also made participants significantly more likely to complete the tasks for four of the five numerical-response task types: \textit{Estimate Administrative Unit}, both \textit{Compare} task types, and \textit{Detect Change in Administrative Unit}.

For tasks involving selectable legends and grid lines, the smallest grid size was selected for 60.9\% of the cartograms as participants entered their responses; 15.9\% of the answers were given while the medium grid size was displayed, and 23.5\% of the questions were answered while the largest grid size was visible.
The smallest grid size was the default setting, which explains why the smallest grid size was most frequently displayed.
Fisher's exact test, applied to a contingency table of correctness versus grid size, resulted in a $p$-value of 0.80.
Thus, there is no evidence that the proportion of correct answers depended on the chosen grid sizes.

Overall, although this treatment was not effective in improving the accuracy of responses and even caused participants to be slower in performing the tasks, it was effective in increasing the task completion rate. 
Participants rated the selectable legend with grid lines to be the most helpful (4.64 out of 5) and most essential (4.45 out of 5) among the three additional features (i.e., legend, grid lines, and selectable legend with grid lines).

\subsection{Effect of the Number of Administrative Units and Individual Traits of Participants on Response Times}

One may suspect that a larger number of administrative units resulted in longer response times because it took longer to discover them on a map.
It is also plausible that the response times depended on individual traits of participants (e.g., confidence in their own map reading skills).
To determine whether these hypotheses are correct, we conducted model selection based on the Bayesian information criterion for linear regression models with and without random effects caused by individual traits of participants (see Section 11 in the supplemental text, available online).
In the base model, we included fixed effects for task type, the number of administrative units, and the treatment (i.e., available features).
In other models, we included fixed effects for participants' responses to the preliminary questions about their familiarity with maps (see Section~\ref{sec:procedure_and_design}).
The input consisted of all correct and non-missing responses.
For \textit{Compare} and \textit{Detect Change} tasks, a response was treated as correct if the direction of change (smaller versus larger) was detected correctly and the participant gave a numeric estimate, regardless of the magnitude of that estimate.

The model with smallest Bayesian information criterion included potential differences between participants as a random effect.
According to this model, the response time increased by approximately 0.9\% for every additional administrative unit.
The effect is statistically significant; the $p$-value of a one-sided $t$-test is $< 10^{-9}$.
The effect was also practically significant because, from 3 units (regions in Belgium) to 49 units (states in the conterminous United States plus Washington, D.C.), the model predicts a 53\% increase in response time.
There are two possible ways to explain why the traits of individual participants resulted in a substantially lower Bayesian information criterion than accounting only for their self-reported map familiarity.
Either the preliminary questions did not effectively measure map familiarity; or other individual traits (e.g., motivation or meticulousness) played a greater role than map familiarity.
Thus, there is no evidence that prior familiarity with the depicted countries influenced the results.

Although the number of administrative units affected response times, it was not a confounding factor.
The Latin-square design ensured that each map was used exactly once in the four different treatments for each task type.
The map and, consequently, the number of administrative units for each treatment was randomly assigned before the experiment (see Section 1 in the supplemental text, available online).

\subsection{Recommendations}
The experimental results suggest that the three additional features that we tested (i.e., static legend, static grid lines, and selectable legend with grid lines) did not affect how accurately the participants performed the seven task types.
At the same time, there is some indication that the additional features negatively affected the participants' speed.
However, we saw a significant increase in the task completion rate when participants had access to any of the three features, especially the selectable legend with grid lines, which reduced the percentage of missing responses to less than 12\% for all the task types that required numerical responses.
There are advantages and disadvantages to adding grid lines or a selectable legend.
Thus, we believe that the decision to include any of the features in a cartogram should depend on the individual use case.

\subsubsection{Cartograms Displayed in a Web Browser}

\begin{figure*}[!tp]
  \centering
    \includegraphics[width=0.8\textwidth]{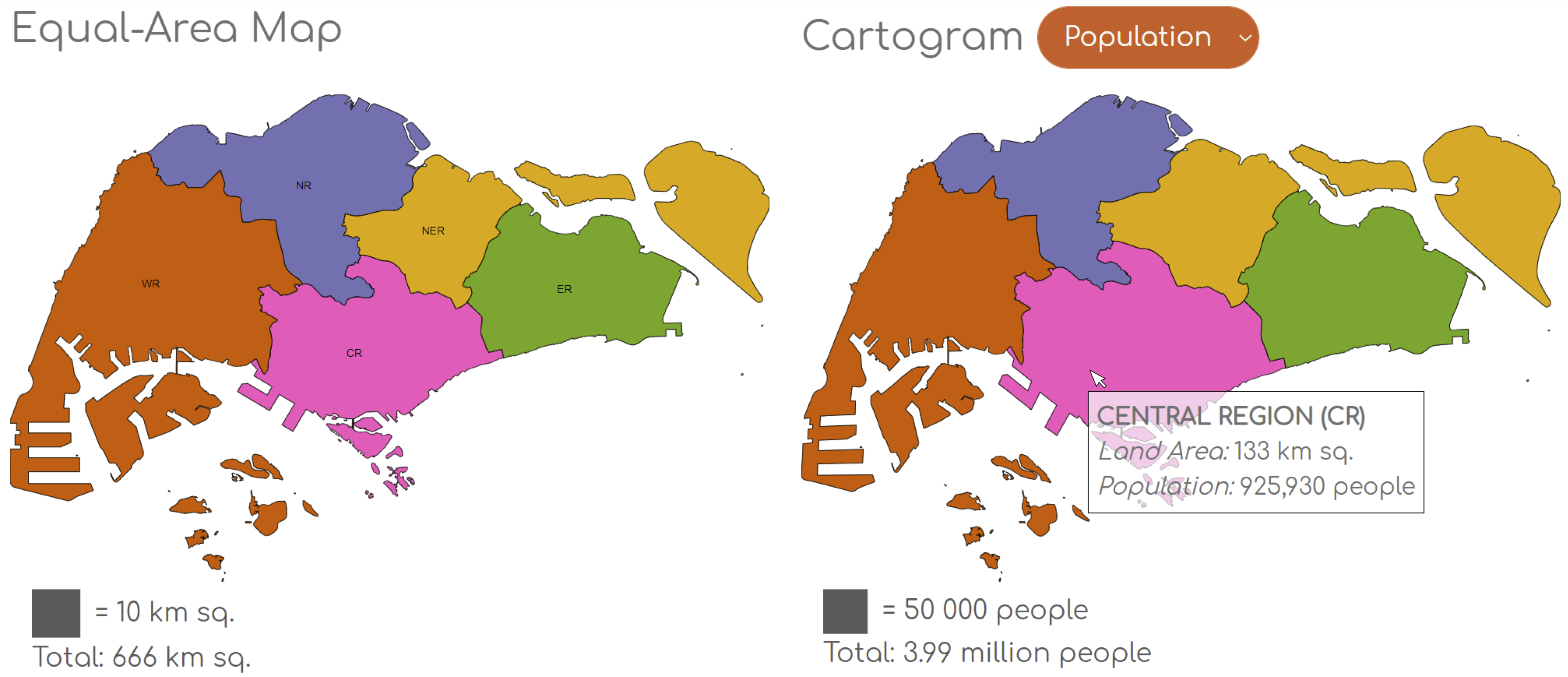}
  \caption{Example of the infotip feature shown on maps of Singapore.
  When the user hovers the mouse cursor over a region, a pop-up appears that contains the region's name and numerical data.
  Screenshot from \href{https://go-cart.io}{go-cart.io}~\cite{gastner_creating_2021}.}
  \label{fig:infotip}
\end{figure*}

In web-based cartograms, it is possible to include more interactive features than those investigated in the present study.
For example, an infotip~\cite{duncan_task-based_2021} (i.e., a text box with information about the region at the position of the mouse pointer) serves a similar purpose as legends and grid lines; that is, it allows participants to infer the magnitude of the numerical data value that is represented by a region in a cartogram. 
As illustrated in Fig.~\ref{fig:infotip}, an infotip can show precise values. 
Consequently, it is possible to retrieve information more accurately and quickly with an infotip than with legends or grid lines.
However, we still recommend including a static legend by default because it communicates the magnitude of the displayed mapping variable~\cite{tingsheng_motivating_2020}. 

Unlike a static legend, we do not recommend including grid lines or a selectable legend by default.
Although the participants rated the selectable legend with grid lines to be helpful rather than hindering, we believe that not all users of web-based cartograms intend to perform tasks that are similar to those in our experiment. 
Moreover, the grid lines and the nested design of the selectable legend add a substantial amount of non-data ink.
Therefore, we recommend a toggle for this feature to be added to web-based cartograms so that users can decide for themselves whether they want to activate it.
A button to toggle grid lines on or off would be straightforward to use, and the button would only require little extra space in the graphical user interface.

\subsubsection{Cartograms in Slide Shows or Video Presentations}

Common slide show and video formats (e.g., PowerPoint and MP4) do not allow individual access to interactive graphics during a presentation.
As for any other medium, we recommend including a legend because it contains essential information that allows relating areas to quantitative data.
However, for a slide show or video, there is no need for the legend to be selectable.

Here, the intended purpose of the cartogram should be the main factor that determines whether static grid lines should be added.
If the cartogram is intended solely for its visual impact, then adding grid lines may not be necessary.
Another consideration is the amount of time the viewers are given to look at the cartogram.
On the one hand, a teacher may include a cartogram in a PowerPoint presentation and give students enough time to extract numerical information from the cartogram.
On the other hand, a business consultant may include a cartogram in a PowerPoint presentation and may only display the slide for a few seconds.
In these scenarios, the teacher should show a cartogram with grid lines, whereas the consultant should use a cartogram without them.

\subsubsection{Cartograms Printed on Paper}

If the cartogram is to be printed on paper, it is impossible to include interactive features such as a selectable legend or an infotip.
In this case, we recommend adding a static legend with grid lines.
Although our results do not indicate that the grid lines improved accuracy, they increased the number of participants completing the tasks.
Therefore, readers are more likely to engage with a cartogram, rather than quickly passing over it, when grid lines are present.

\section{Conclusion}

Cartograms are a useful type of data visualization because they simultaneously show geography and statistics.
However, previous studies have shown that retrieving quantitative information from cartograms is not a trivial task~\cite{dent_communication_1975, griffin_recognition_1983, kaspar_empirical_2011}.
The purpose of this study was to find out whether legends and grid lines, both with and without interactivity, can support information retrieval from contiguous area cartograms.
The results of our experiment show that these additional features cause map readers to be slower in estimating numerical values.
The estimates are less variable, but not more accurate, when legends or grid lines are added.
However, the additional features, especially the selectable legend with grid lines, have the positive effect that they significantly increase the map readers' confidence so that readers are more likely to complete cartogram reading tasks.
This study was limited in scope to contiguous cartograms.
For rectangular cartograms~\cite{van_kreveld_rectangular_2004}, grid lines would probably have fewer benefits because the boundaries are already horizontally and vertically aligned.
For mosaic cartograms~\cite{cano_mosaic_2015}, in which equally sized hexagonal or square tiles imply a fixed unit, grid lines are superfluous as long as individual tiles are discernible (e.g., by indicating their borders with thin lines).

Thus, we do not recommend incorporating grid lines into cartograms as a standard practice. 
Instead, we believe that it is crucial to examine the use case and the intended function of the cartogram before deciding which additional features are to be included.
For contiguous cartograms, our study provided quantitative evidence that legends and grid lines make readers more confident about their assessment of the presented data.
We hope that our recommendations help cartograms to achieve their potential to become ``a more socially just form of mapping''~\cite[p.~4]{dorling_area_1996} that effectively highlights inequalities (e.g., in income~\cite{cruz_adapted_2017} and health~\cite{yau_mapping_2021}) to a wide audience.


%



\ifCLASSOPTIONcompsoc
  \section*{Acknowledgments}
\else
  \section*{Acknowledgment}
\fi
We thank Ian~K.~Duncan and Chen-Chieh Feng for helpful discussions.
We would like to acknowledge Ozair Faisal's assistance in supervising the experiment participants.
We are indebted to the editor and three anonymous referees for their insightful comments about an early draft of this manuscript.
This research is supported by the Ministry of Education, Singapore, under its Academic Research Fund Tier 2 programme (award number MOE-T2EP20221-0007) and its Academic Research Fund Tier 1 programme (Grant IG18-PRB104, R-607-000-401-114), and by Yale-NUS under its Summer Research Programme.
We would like to thank Editage (www.editage.com) for English language editing.

\ifCLASSOPTIONcaptionsoff
  \newpage
\fi



%

\bibliographystyle{IEEEtran}
\bibliography{cartogram_legend}

%

\begin{IEEEbiography}[{\includegraphics[width=1in,height=1.25in,clip]{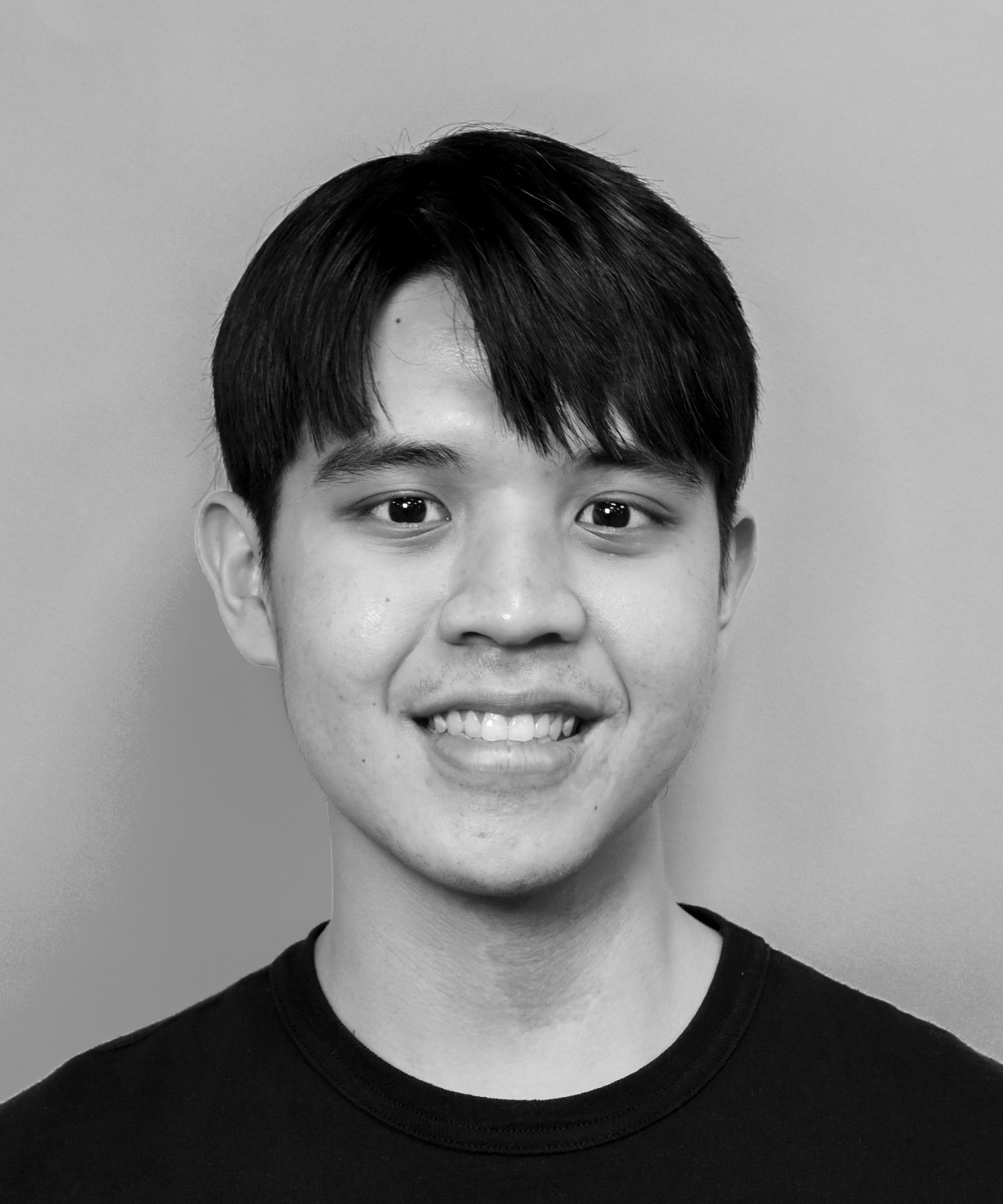}}]{Kelvin~L.T.~Fung} received his  B.Sc.\ in mathematical, computational, and statistical sciences from Yale-NUS College (Singapore) in 2021 and his M.Sc.\ in machine learning from University College London in 2022.
\end{IEEEbiography}

\begin{IEEEbiography}[{\includegraphics[width=1in,clip,keepaspectratio]{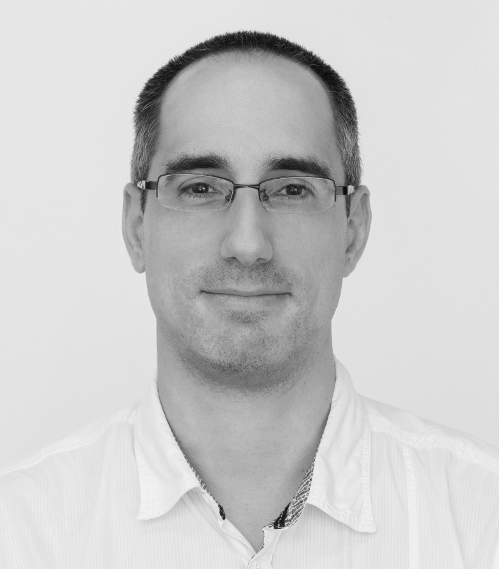}}]{Simon T.~Perrault} obtained his PhD degree in computer science from Telecom ParisTech (France) in 2013.
He is an Assistant Professor in Information Systems Technology and Design at the Singapore University of Technology and Design (SUTD).
Previously, he was a Visiting Professor at the Korean Advanced Institute of Science and Technology (Korea), and Assistant Professor at Yale-NUS College (Singapore).
\end{IEEEbiography}

\begin{IEEEbiography}[{\includegraphics[width=1in,height=1.25in,clip,keepaspectratio]{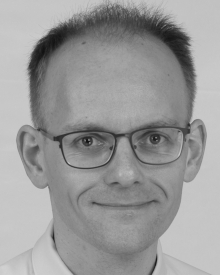}}]{Michael~T.~Gastner} received his PhD in physics from the University of Michigan in 2005.
He is an Associate Professor in Infocomm Technology at the Singapore Institute of Technology.
Previously, he held academic positions at the Santa Fe Institute, the University of Oldenburg, Imperial College London, the University of Bristol, the Hungarian Academy of Sciences, and Yale-NUS College.
\end{IEEEbiography}





\end{document}